\documentclass[10pt,a4paper]{article}

\usepackage{indentfirst}			
 \usepackage{enumerate}				
\usepackage{amsmath, amsgen,amsfonts,amssymb,amsbsy}	% Enlarge math symbols set

\def\CQG{\it Class.\ Quantum Grav.\ }
\def\PRD{\it Phys.\ Rev.\ D\ }

\begin{document}

\thispagestyle{plain}		% Remove headings in first page

\title{Vacuum Static Spherically Symmetric Spacetimes in Harada's Theory}
\author{Alan Barnes\\ 
26 Havannah Lane, \\
Congleton CW12 2EA, \\
United Kingdom. \\
E-Mail: \ \ {\tt Alan.Barnes45678{\bf @}gmail.com}}
\maketitle

\begin{abstract}
\noindent Very recently Harada proposed a gravitational theory which is of third
order in the derivatives of the metric tensor with the property that any solution
of Einstein's field equations (EFEs) possibly with a cosmological constant
is necessarily a solution of the new theory. He then applied his theory
to derive a second-order ODE for the evolution of the scale factor of the FLRW
metric. Remarkably he showed that, even in a matter-dominated universe with zero
cosmological constant, there is a late-time transition from decelerating to
accelerating expansion.
Harada also derived a generalisation of the Schwarzschild solution.
However, as his starting point he assumed an
unnecessarily restricted form for a static spherically symmetric metric. In this
note the most general spherically symmetric static vacuum solution of the theory
is derived.

Mantica and Molinari have shown that Harada's theory may be recast into
the form of the EFEs with an additional source term in the form of a
second-order conformal Killing tensor(CKT).  Accordingly they have dubbed the theory
\emph{conformal Killing gravity}. Then, using a result in a previous paper of theirs
on CKTs in generalised Robertson-Walker spacetimes, they rederived Harada's
generalised evolution equation for the scale factor of the FLRW metric.

However, Mantica and Molinari appear to have overlooked the fact that all
solutions of the new theory (except those satisfying the EFEs) admit a non-trivial
second-order \emph{Killing} tensor. Such Killing tensors are invaluable when
considering the geodesics of a metric as they lead to a second quadratic invariant
of the motion in addition to that derived from the metric.
\end{abstract}

\vspace{-5 pt}
\section{Introduction}
\noindent Recently Harada\cite{harada} has proposed a new gravitational theory
satisfying three theoretical criteria for generalised theories of gravity namely:
\begin{description} 
\item The cosmological constant $\Lambda$ is obtained as a constant of integration.
\item The conservation law $T^a_{b;a}=0$ is a consequence of the field equations.
\item Conformally flat metrics are not necessarily vacuum.
\end{description}

Applying the above criteria he was led to consider the totally symmetric derivatives
of a trace-modified  Einstein tensor $\tilde{G}_{ab}$
\begin{equation}
  H_{abc} = \tilde{G}_{(ab;c)}\quad\mathrm{where} \quad
  \tilde{G}_{ab} = R_{ab}-\frac{1}{3}Rg_{ab}=   G_{ab}-\frac{1}{6}Gg_{ab}
  \label{tildeG}
\end{equation}
where round brackets indicate symmetrisation; and the similarly modified
energy-momentum tensor
\begin{equation}
  T_{abc} = \tilde{T}_{(ab;c)} \quad\mathrm{where} \quad
  \tilde{T}_{ab} = T_{ab}-\frac{1}{6}Tg_{ab}
  \label{tildeT}
\end{equation}
and to adopt as the field equations of his theory:
\begin{equation}
  H_{abc} = T_{abc}.
  \label{hfe}
\end{equation}
The  vacuum case in this theory is characterised by the condition $T_{abc} =0$.

The energy-momentum conservation equation follows from \eqref{hfe} by contraction:
\[ g^{ac}H_{abc}=G^a_{b;a} = 0 = g^{ac}T_{abc} =T^a_{b;a}. \]
It also follows immediately  that any solution of the EFEs: $G_{ab}=T_{ab}$ automatically
satisfies Harada's field equations \eqref{hfe}. A similar conclusion holds for solutions
of the EFEs with a cosmological constant $G_{ab}+\Lambda g_{ab}=T_{ab}$.

Harada\cite{harada} went on to consider the evolution of the scale factor $a(t)$
of the Friedmann-Lema\^{i}tre-Robertson-Walker metric:
\begin{equation}
ds^2 = \mathrm{d}t^2 - a^2(t)\left(\frac{\mathrm{d}r^2}{1-kr^2} + r^2\mathrm{d}\theta^2 +
  r^2\sin^2\theta \mathrm{d}\phi^2\right)
  \label{FLRW}
\end{equation}
in his theory and obtained a third order ODE which has the first integral:
\begin{equation}
  2\left(\frac{\dot{a}}{a}\right)^2-\frac{\ddot{a}}{a}+\frac{2k}{a^2}=
    \frac{4\pi G}{3}(5\rho +3p)+\frac{\Lambda}{3}.
 \label{evoleq}
\end{equation}
He went on to show that, even in the case of a matter-dominated universe ($p=0$)
with $\Lambda=0$, there was a transition to accelerating expansion. In a
second paper Harada\cite{harad2} considered this problem in greater depth and
suggested that his theory also had the potential to address the Hubble tension
problem.

Mantica and Molinari\cite{mantmol} have examined Harada's field equations and
shown that they can be recast in the form of Einstein's field equations with an
additional source term which is a second-order conformal Killing tensor $C_{ab}$
defined by $C_{ab} = G_{ab}-T_{ab}$. The trace $C$ of $C_{ab}$ is given by $C=G-T$
and a straightforward calculation using
\eqref{hfe} shows that
\begin{equation}
  C_{(ab;c)} = \frac{1}{6}g_{(ab}C_{,c)},
  \label{CKT}
\end{equation}
and hence $C_{ab}$ is a gradient conformal Killing tensor. Hence \eqref{hfe}
is equivalent to the `Einstein' equation
\[ G_{ab}=T_{ab}+C_{ab}. \]
Earlier they\cite{mantmol2} had investigated generalised Robertson-Walker
spacetimes and shown that they must admit a gradient conformal Killing tensor of
the form $C_{ab} = Bu_au_b+Ag_{ab}$ where $A$ and $B$ are scalar fields and $u_a$
is the velocity vector. They were then able to use this result to rederive
Harada's first integral \eqref{evoleq} of the evolution equations for the scale
factor $a(t)$ and to independently obtain the results of \cite{harad2}.

\section{Static Spherically Symmetric Vacuum Solutions}
\noindent Harada\cite{harada} derived the following exact solution of his
field equations
\eqref{hfe} for the static spherically symmetric vacuum case:
\begin{equation}
  \mathrm{d}s^2 = e^{2a(r)}\mathrm{d}t^2 - e^{-2a(r)}\mathrm{d}r^2 -
      r^2(\mathrm{d}\theta^2+\sin^2\theta\mathrm{d}\phi^2),
   \label{sssmet}
\end{equation}   
where \begin{equation}
  e^{2a(r)} = 1-2m/r-\Lambda r^2/3-\lambda r^4.
  \label{schwlike}
\end{equation}
$m$, $\Lambda$ and $\lambda$ are arbitrary constants of integration. When
$\lambda =0$, this is the well-known Schwarzschild-de Sitter metric, but the
$\lambda r^4$ term is a new feature of the theory.

However, as his starting point Harada assumed the metric had the form
\eqref{sssmet} which is not the most general form. In terms of curvature
coordinates the most general spherically symmetric static metric is
\begin{equation}
  \mathrm{d}s^2 = e^{2a}\mathrm{d}t^2 - e^{2b}\mathrm{d}r^2 -
      r^2(\mathrm{d}\theta^2+\sin^2\theta\mathrm{d}\phi^2),
   \label{gensssmet}
\end{equation}
where $a$ and $b$ are functions of $r$ only.

The computer algebra system Classi (see \cite{Aman} and \cite{ACGR}) was used to
calculate the components of the tensor $H_{abc}$.
In terms of the obvious Lorentz orthonormal tetrad of one forms:
\begin{equation}
  e^a\mathrm{d}t,\qquad e^b\mathrm{d}r,\qquad r\mathrm{d}\theta, \qquad
  r\sin\theta\mathrm{d}\phi,
\label{tetrad}
\end{equation}
the only non-zero \emph{frame} componentents of $H_{abc}$ are $H_{rtt}$,
$H_{rrr}$ and $H_{r\theta\theta}=H_{r\phi\phi}$. In fact only two of these are
linearly independent as $2H_{r\theta\theta}-H_{rtt}+H_{rrr} \equiv 0$. It is
convenient to work with the two equations:
\begin{equation}
  3H_{rtt}+H_{rrr} = r(a''-2a'^2-4a'b'+b''-2b'^2)-a'-b' =0
  \label{eqn1}
\end{equation}
and
\begin{eqnarray}
  H_{rtt} &=& -r^3(a'''+2a''a'-3a''b'-2a'^2b'-a'b''+2a'b'^2)\nonumber\\ 
  & & + r^2(4a''-8a'b'+2b''-4b'^2)-r(4a'+6b')+4e^{2b} -4 =0,
  \label{eqn2}
\end{eqnarray}
where a prime denotes differentiation with wrt $r$.

Using the substitution $b=f-a$ and cancelling common factors \eqref{eqn1}
simplifies to
\begin{equation}
  r f''-2r f'^2 -f' =0.
  \label{feqn}
\end{equation}
This equation may be integrated to yield
\begin{equation}
  f=-log(c+dr^2)/2
  \label{genf}
\end{equation}
where $c$ and $d$ are arbitrary constants of integration.
Hence the metric takes the form \eqref{gensssmet} with
$e^{2b} = e^{-2a}/(c+dr^2)$.  Eliminating $b$ from \eqref{eqn2} and removing
common factors we obtain
\begin{eqnarray}
  \lefteqn{(r^3(c+dr^2)(a'''+6a'a''+4a'^3)}\nonumber\\
  & &-r^2(2c-dr^2)(a''+2a'^2) -r(2c+dr^2)a'+4(c-1) =0.
\end{eqnarray}
By applying the substitution $y=e^{2a}$, this equation may be simplified to
produce the linear equation
\begin{equation}
 (c+dr^2)r^3y''' - (2c-dr^2)r^2y''-(2c+dr^2)ry'+ 8cy = 8
\label{gensol}
\end{equation}
which may be solved by using standard textbook methods.

If $d=0$, then $e^b= e^{-a}/c$  and after a suitable constant rescaling of the
$t$ coordinate, $c$ may be set to 1. Equation \eqref{gensol} reduces to
\begin{equation}
  r^3y'''-2r^2y''-2ry'+8y = 8.
\end{equation}
This is easily integrated to yield
$y= e^{2a} = 1-2m/r-\Lambda r^2/3-\lambda r^4/5$ where
$m$, $\Lambda$ and $\lambda$ are arbitrary constants of integration. Following
Harada the numerical factors of the $1/r$ and $r^2$ terms have been chosen to
correspond with those of the  Schwarzschild-de Sitter solution.
Thus Harada's solution in \eqref{sssmet} and \eqref{schwlike} is
obtained.

If  $c=0$, a similar rescaling of the $t$ coordinate may be used to set
$d=1$ and $e^{2b} = e^{-2a}/r^2$. Equation \eqref{gensol} now reduces to
\begin{equation}
  r^5y'''+r^4y''-r^3y' = 8.
\end{equation}
This may be integrated to yield
\begin{equation}
  y =e^{2a} = \lambda -\Lambda r^2/3 + m \log r -1/(2r^2)\qquad
  e^{2b} = e^{-2a}/r^2
  \label{logsol}
\end{equation}
where $m$, $\Lambda$ and $\lambda$ are again arbitrary constants of
integration.

In the general case where both $c$ and $d$ are both non-zero, $c$ may
be again set to 1 by a rescaling of the $t$ coordinate. In this case
the solution of \eqref{gensol} involves infinite power series and is
obtained by  use of the well-known Frobenius method. Firstly there is an
obvious \emph{particular integral} $y=1$. The \emph{complementary function}
is obtained by searching for solutions of the form
\begin{equation}
  y=\sum_{n=0}^\infty a_nr^{c+n}.
\end{equation}
and equating the coefficient of each power of $r$ in \eqref{gensol}
to zero.  For $n=0$ this results in the \emph{indicial equation}
$c^3-5c^2+2c+8 = (c-2)(c-4)(c+1) =0$.
The case $c=2$ leads to the monomial solution $y=r^2$ (the cosmological
constant term). The cases $c=4$ and $c=-1$ each result in an infinite power
series; for $n=1$ we obtain $(c^3-2c^2-5c+6)a_1=0$ and thus $a_1=0$ in both
cases. For  $n>=2$ and $c=-1$ the recurrence relation $a_n=
-d\frac{n-3}{n}a_{n-2}$ is obtained whilst for $c=4$ we obtain
$a_n=-d\frac{n+2}{n+5}a_{n-2}$.  Clearly in both cases the coefficient $a_n$
vanishes when $n$ is odd. Also the radius of convergence of both series is
easily seen to be $1/\sqrt{|d|}$.
The general solution of \eqref{gensol} is therefore
\begin{eqnarray}
  y = e^{2a} &=& 1 -2mp(r)/r +\lambda q(r)r^4/5 -\Lambda r^2/3\\
  \mathrm{where} \quad p(r)&=& 1 +dr^2/2 -d^2r^4/8+d^3r^6/16\nonumber\\
  & & -5d^4r^8/128+7d^5r^{10}/256\ldots\\
  \mathrm{and}\quad q(r) &=& 1 -4dr^2/7 + 8d^2r^4/21-64d^3r^6/231\nonumber\\
  & & +640d^4r^8/3003-512d^5r^{10}/3003  \ldots \\
  \mathrm{with} \quad e^{2b} &=& e^{-2a}/(1+dr^2).
\end{eqnarray}
where $m$, $\Lambda$ and $\lambda$ are again arbitrary constants of integration.

If we set $m=0$ and $\lambda=0$ in the general solution above, the following
metric is obtained
\begin{equation}
  \mathrm{d}s^2 = (1-\Lambda r^2/3)\mathrm{d}t^2 -
  \frac{1}{(1-\Lambda r^2/3)(1+d r^2)}\mathrm{d}r^2 -
      r^2(\mathrm{d}\theta^2+\sin^2\theta\mathrm{d}\phi^2).
\label{specsol}
\end{equation}

As all the solutions obtained in this section are spherically symmetric they
must be of Petrov type D or conformally flat; in fact apart from the Minkowski metric and de Sitter metric which is a special case of the Harada metric
\eqref{schwlike} with $m=\lambda=0$, the only conformally flat solution is a
special case of \eqref{specsol} with $\Lambda=0$. This is the metric of
Einstein's static universe which is a vacuum solution in Harada's theory!

The only non-zero \emph{frame} components of the Weyl tensor are
\begin{equation}
  C_{0101}=-C_{2323} = A,\qquad C_{1212}=C_{1313} = -C_{0202}=-C_{0303}= A/2,
\end{equation}
where
\begin{equation}
  A = -\frac{r^2(c+dr^2)y''-r(2c+dr^2)y'+2cy-2}{6r^2}.
\end{equation}

\section{Killing Tensors}
\noindent As pointed out by Mantica and Molinari\cite{mantmol},
$C_{ab}=G_{ab}-T_{ab}$ is a gradient conformal Killing tensor (see \eqref{CKT}
in the Introduction). However, a gradient CKT is associated with a Killing
tensor. For if $C_{(ab;c)} = g_{(ab}\eta_{,c)}$ for some scalar field $\eta$, then
$K_{ab} -\eta g_{ab}$ is a \emph{Killing tensor}(see, for example \cite{Rani}).
In fact from \eqref{tildeG}, \eqref{tildeT} and \eqref{hfe}, it is obvious that
$K_{ab} = \tilde{G}_{ab}- \tilde{T}_{ab}$ is a Killing tensor.
If a solution of \eqref{hfe} is also a solution of the EFEs, this
Killing tensor is identically zero whilst for solutions of the EFEs
with a cosmological contant, the Killing tensor is a constant
multiple of the metric tensor. However, for other solutions of
Harada's theory the Killing tensor is non-trivial.

As is well-known (for example \cite{Wald}), a Killing tensor corresponds to a
constant of motion on geodesics; specifically for a geodesic with tangent
vector $u^a$, $K_{ab}u^au^b$ is a constant along geodesics. Since the
trace-modified Eistein tensor $\tilde{G}_{ab}=G_{ab}-\frac{G}{6}g_{ab}$ is a
Killing tensor, the analysis of geodesics in vacuum Harada fields is
considerably simplified by the existence of this second quadratic
constant of motion in addition to that associated with the
metric $g_{ab}u^au^b$.

For the general static spherically symmetric metric \eqref{sssmet} derived in
the previous section the Killing tensor has the following non-zero
\emph{coordinate} components:
\begin{eqnarray}
  K_{tt} & = & y\frac{r^2(c+dr^2)y''-r(2c+dr^2)y'-4(c+3dr^2)y+4}{6r^2}\\
  K_{rr} & = & -\frac{r^2(c+dr^2)y''-r(2c+dr^2)y'-4cy+4}{6r^2y(c+dr^2)}\\
  K_{\theta\theta}  & = & \frac{r^2(c+dr^2)y''+r(c+2dr^2)y'-cy+1}{3}\\
  K_{\phi\phi}  & = & \frac{r^2(c+dr^2)y''+r(c+2dr^2)y'-cy+1}{3}\sin^2 \theta.
\end{eqnarray}
where $y=e^{2a}$ and $e^{2b}= e^{-2a}/(c+dr^2)$. Thus for the geodesics
of the solutions
discussed in section 2, there are four constants of motion:
\begin{equation}
  g_{ab}\dot{x}^a\dot{x}^b,\qquad K_{ab}\dot{x}^a\dot{x}^b,\qquad \xi_a\dot{x}^a,
  \qquad\eta_a\dot{x}^a,
\end {equation}
where $\dot{x}^a$ is the tangent vector of the affinely-parameterised
geodesic and $\xi=\partial_t$ and $\eta=\partial_\phi$ are the
timelike and rotational Killing vectors. Further simplification is
possible since, by a suitable rotation of the $\theta$ and $\phi$
coordinates, the geodesic motion can be chosen to lie in the
`equatorial' plane so that $\theta=\pi/2$ and $\dot{\theta}=0$.

\section{Conclusions}
\noindent All static spherically symmetric vacuum solutions in Harada's `conformal
Killing' gravity theory are derived. The most general solution involves five
parameters; however only the ratio of the fourth and fifth parameters $c/d$ is
essential; either $c$ or $d$ (unless zero) may be set to unity by a constant
rescaling of the $t$ coordinate.

The general solution when $c$ and $d$ are both non-zero involves two
infinite power series in the curvature coordinate $r$ plus a `cosmological constant'
term which is a multiple of $r^2$.
Harada's original three-parameter `Schwarzschild-like' solution is obtained when the
parameter $d=0$; the two power series degenerating to monomials in this case. Two of
the  parameters may be identified as the mass $m$ and the cosmological constant
$\Lambda$, but the third $\lambda$ and the ratio $c/d$ are new and have no analogue
in General Relativity. A second three-parameter solution not involving power series
is obtained when $c=0$; its physical interpretation is currently unclear.

All solutions of Harada's theory other than those satisfying the Einstein field
equations are shown to admit a non-trivial second-order Killing tensor. The quadratic
constant of motion associated with this Killing tensor simplifies the analysis
of geodesic motion in Harada gravitational fields especially the static spherically
symmetric vacuum fields derived in section 2.

A number of important questions arise regarding the theory. For example, are the
spherically symmetric vacuum solutions of the theory necessarily static as they are
in General Relativity. If not, then it would appear that a spherically symmetric
collapsing or pulsating star ought to generate gravitational waves leading to
possible  experimental evidence regarding the theory.  In this context a study of
the propagation of gravitational waves in the weak field approximation and/or a
study of exact plane wave solutions would be useful.

There are a number of perhaps interesting theoretical questions; for example what are
the asymptotic and event horizon structures of the solutions derived in section 2.
There is clearly a curvature singularity at $r=0$ except when $m=0$ and if either
$\Lambda$ or $\lambda$ is non-zero, the solutions are not asymptotically flat.

Senovilla\cite{seno} considered junction conditions in $\mathrm{F}(R)$ gravity
which is fourth order in the metric derivatives. He showed that interesting new
features can arise such as gravtational double layers. It would be interesting to see
whether such features appear in Harada's theory which is third order.

\section*{Acknowledgements}
\noindent The extensive calculations in sections 2 and 3 were performed using  the
Sheep/Classi package for General Relativity which was kindly supplied to me by Jan
{\AA}man of the University of Stockholm. I would also like to thank him for useful
discussions on some undocumented features of the system. Some calculations were
also performed using the Reduce computer algebra freely available for download
from SourceForge (https://sourceforge.net/projects/reduce-algebra/files/).
Two source files \texttt{harada.shp} and \texttt{harada.lor} not in the
standard Classi distribution are available from the author on request.


\begin{thebibliography}{99}
\setlength{\itemsep}{-0.6 ex}	% Adjust inter-item separations
\vspace{-5 pt}

\bibitem{harada} J.\ Harada, {\sl Gravity at cosmological distances: Explaining the
accelarating expansion without dark energy\/}, \PRD (to be published),
  arXiv:2308.02115 [gr-qc].

\bibitem{harad2} J.\ Harada, {\sl Dark energy in conformal Killing gravity\/},
Unpublished, arXiv:2308.07634 [gr-qc].

\bibitem{mantmol} J.C.\ Mantica \& L.G.\ Molinari, {\sl A note on Harada's
  Conformal Killing Gravity\/}, Unpublished,  arXiv:2308.06803 [gr-qc].  

\bibitem{mantmol2} J.C.\ Mantica \& L.G.\ Molinari, {\sl Generalized
  Robertson-Walker spacetimes, a survey\/}, \it{Int. J. Geom. Meth. Mod.
  Phys.\/} {\bf 14} n.3, 1730001 (2017) arXiv:1612.07021 [gr-qc]

\bibitem{Aman}  J.\ E.\ {\AA}man, {\sl Classification programs for geometries in general relativity -- manual for CLASSI, 4th edition, report Univ.\ Stockholm} (2002).

\bibitem{ACGR} M.A.H.\ MacCallum \& J.E.F.\ Skea, {\sl SHEEP: A computer
  algebra system for general relativity\/}, \emph{Chapts.\ 1--6 in:}
  {\sl Algebraic Computing in General Relativity\/},
  Oxford University Press (1994).

\bibitem{Rani} R.\ Rani, S.B.\ Edgar \& A.\ Barnes, {\sl Killing tensors and
  conformal Killing tensors from conformal Killing vectors\/} \CQG {\bf 20}
  (11), 1929 (2003),  arXiv:0301059 [gr-qc].

\bibitem{Wald} R.M.\ Wald, {\sl General Relativity\/}  p.444, University
  of Chicago Press (1984).

\bibitem{seno} J.M.M.\ Senovilla, {\sl Junction conditions for F(R)-gravity and
  their consequences\/} \PRD {\bf 88}, 064015 (2013), arXiv:1303.1408 [gr-qc]. 

\end{thebibliography}
\end{document}